# Room temperature green light emissions from nonpolar cubic InGaN/GaN multi quantum wells

Shunfeng Li *, Jörg Schörmann, Donat. J. As, Klaus Lischka [a)]

Universität Paderborn, Department of Physics, D-33098 Paderborn, Germany.

**Abstract**

Cubic InGaN/GaN multi quantum wells (MQWs) with high structural and optical quality are achieved by utilizing free-standing 3C-SiC (001) substrates and optimizing InGaN quantum well growth. Superlattice peaks up to $5^{th}$ order are clearly resolved in X-ray diffraction. We observe bright green room temperature photoluminescence (PL) from c-$In_xGa_{1-x}N$/GaN MQWs (x=0.16). The full-width at half maximum of the PL emission is about 240 meV at 300 K. The PL intensity increases with well thickness, proofing that polarization fields which can limit the performance of the wurtzite III-nitride based devices are absent. The diffusion length of excess carriers is about 17 nm.

Electronic mail: lischka@upb.de

Group III-nitrides crystallize in the stable wurtzite (hexagonal) structure or in the metastable zincblende (cubic) structure. Currently, state-of-art nitride-based devices are grown along the polar c-axis of the wurtzite III-nitride unit cell. Due to strong polarization field along the c-axis, the electron-hole pairs within quantum wells are separated, which results in a red-shift of optical transitions and a reduced oscillator strength. This so called quantum confined Stark effect (QCSE) [1,2] can reduce the efficiency of optoelectronic devices which employ InGaN/GaN quantum wells in the active region. Thus in highly efficient wurtzite InGaN/GaN multi quantum well (MQW) based devices, the InGaN well has a thickness of less than 3 nm to avoid degradation of the light emission by the QCSE [3]. Polarization field effects can also be detrimental for electronic devices, e.g. normally off field effect transistors [4].

To overcome these problems wurtzite GaN grown along non-polar direction, e.g. a- and m- directions have been explored by several groups [5-7]. Recent results reveal the absence of polarization induced electric field in a-plane and m-plane grown wurtzite III-nitride MQWs. However, the structural and optical properties of nonpolar wurtzite III-nitrides are still inferior compared to their mature c-plane grown counterpart.

Due to the centro-symmetric unit cell of cubic III-nitrides polarization and piezoelectric fields are absent when they are grown in an (001) direction. Furthermore, due to the relatively narrower bandgap (about 200 meV) of cubic III-nitrides, InGaN layers containing less In can be used to extend the light emission to the green spectral range, which is still a challenge for high efficiency wurtzite III-nitride based optoelectronic devices.

In this paper we show that by optimizing the growth conditions and employing free-standing 3C-SiC (001) substrates, cubic InGaN/GaN MQWs with improved quality can be fabricated. Their optical and structural properties are comparable to that of non-polar (11-20) a-plane wurtzite InGaN/GaN MQWs. The absence of any polarization fields in our c- InGaN/GaN MQWs is demonstrated. We report strong green emission from c- InGaN MQWs with 7.8 nm thick $In_{0.16}Ga_{0.84}N$ well layers.

Samples investigated in this work were grown on free-standing 3C-SiC (001) substrates by molecular beam epitaxy (MBE). The MBE system is equipped with an Oxford RF plasma source for activated nitrogen. Indium and Gallium is evaporated from conventional Knudsen cells and the metal fluxes were adjusted by the source temperatures. Prior to growth, the 3C-SiC substrates were chemically etched and annealed for 10 hours at 500 °C. GaN buffer layers were grown at 720 °C with a growth rate of 100 nm/h under controlled Ga-rich conditions. On top of the c-GaN buffer layers 6-period $In_xGa_{1-x}N$/GaN MQWs were deposited. The InGaN layers were grown at 610 °C under slightly In-rich growth conditions. Indium rich growth helps to improve the interface quality by utilizing the indium surfactant effect [8]. However, excess In flux leads to an enhanced In surface segregation, which is detrimental for the interface quality. Therefore, growth was interrupted after the deposition of each quantum well with the InGaN layer at growth temperature. These interruptions have been proven to remove effectively the surface In atoms and relieve the In segregation effect [9]. In a next step a thin GaN layer was deposited to avoid the decomposition of the InGaN layer during the following ramping-up of the substrate temperature. The remaining GaN barrier layers were grown at a substrate temperature of 720 °C. The InGaN well thickness was varied from 2 to 8 nm, the barrier thickness was about 12.0 nm unless otherwise mentioned.

Photoluminescence (PL) measurements were performed at room temperature and 2 K using excitation by a 325 nm HeCd laser. The light intensity on the samples was about 10W/cm$^2$. The luminescent light was dispersed by a monochromator and detected by a cooled photomultiplier. A Philips X'pert high resolution x-ray diffractometer (HRXRD) was used to characterize the structural properties of the samples. The HRXRD diffraction profiles of our c-InGaN/GaN QWs were analyzed by a dynamic diffraction model which allowed deriving the parameters of the quantum wells.

Figure 1 displays experimental and simulated HRXRD profiles of a typical 6-period c-$In_xGa_{1-x}N$/GaN MQW. Pronounced superlattice satellite peaks up to the fifth order are observed indicating a high structural quality of the GaN/InGaN interfaces. A simulation using dynamic diffraction theory has been performed; the results are also shown in Fig. 1. The calculated curve agrees well with the experimental data, yielding a well thickness of 3.4 nm, an indium mole fraction of 0.16 and a barrier thickness of 14.9 nm, respectively. A reciprocal space map of the (-1-13) Bragg reflex revealed that all InGaN QWs are pseudomorphically grown on the c-GaN buffer.

Room temperature and 2 K PL spectra of a 6-period c-$In_xGa_{1-x}N$/GaN (x=0.12)

MQW are shown in Fig. 2.a and 2.b. The well thickness of this sample is 3.7 nm. Strong quantum well emission at 2.57 eV and 2.65 eV is observed at 300 K and 2 K, respectively. The full-width at half maximum (FWHM) of the InGaN emission are 240 meV and 190 meV for 300 K and 2 K, which compare favorably to the values reported for nonpolar a-plane wurtzite InGaN/GaN MQWs with similar emission energy [5]. The two additional peaks at 3.13 eV and 3.26 eV in Fig. 2a originate from the donor-acceptor pair transition and bound exciton trasition in the GaN barriers, respectively [10].

Quantum wells enhance the light emission by increasing oscillator strength and collecting the carriers generated within an effective diffusion length ($L_{EDL}$) of the barriers. We roughly estimate $L_{EDL}$ by comparing the PL intensity of this MQW sample with that of a single quantum well sample with 4.5 nm thick well and similar In mole fraction. Room temperature PL spectra of both samples were recorded under identical conditions. Since the GaN emission is almost zero in the PL spectrum of MQW sample (see Fig. 2(a)), we conclude that all the electron-hole pairs generated in the barrier layers of the MQWs either diffuse into the InGaN well layers or recombine nonradiatively, revealing that $L_{EDL}$ exceeds the barrier thickness of the MQWs.

Assuming (i) the same carrier generation rates and nonradiative recombination rate in the GaN barriers and the InGaN quantum wells, and (ii) no variation of the excitation intensity in that region, where those electron-hole pairs are excited which contribute to the emission, and (iii) surface recombination can be neglected we can calculate the effective carrier diffusion length $L_{EDL}$ by the following equation:

$$2.6 * (L_{SW} + 2L_{EDL}) = 6 * (L_{MW} + L_B) + L_{EDL} \tag{1}$$

where $L_{SW}$ is the well thickness of the SQW, $L_{MW}$ and $L_B$ are the well and the barrier thickness of MQWs, respectively. The factor of 2.6 on the left hand side is the ratio of the integrated PL intensity from MQWs (Fig. 2(a)) and a SQW (not shown). The barrier thickness of the MQWs sample is 10.3 nm. The thickness of the top layer of the MQW sample is equal to the barrier thickness, therefore six $L_B$ appear on the right hand side of Equ. (1). Using Equ. (1) we obtain an effective diffusion length of excess carriers in our samples of about 17 nm. In our calculation we have neglected the coupling between QWs since numerical calculations using a program of I.H.Tan et al [11] show that it is extremely week for barriers as thick as in our InGaN/GaN MQW samples. We therefore assume an equal recombination rate in the single QW and the individual quantum wells of the MQW structure, respectively.

It has been shown that in h-InGaN/GaN quantum wells grown on c-plane substrates the PL intensity decreases with increasing well thickness due to the separation of electrons and holes by polarization fields [12]. In order to demonstrate the absence of polarization field in our samples, we grew a series of c-InGaN/GaN (x=0.160±0.003) MQWs with identical structural parameters except the quantum well thickness. Room temperature PL spectra of these samples are depicted in Fig. 3. The respective well thickness is indicated in the graphs. The c-InGaN quantum well emissions dominants all spectra. With increasing well thickness a clear red-shift is observed. No emission from the c-GaN barriers is visible, which manifests the high collection efficiency of the InGaN QWs.

Figure 4 depicts the integrated PL intensity versus quantum well thickness. The PL intensity increases monotonically with well thickness up to about 8 nm. This is in clear contrast to what was observed with c-plane h-InGaN/GaN MQWs where the emission intensity decreases with quantum well thickness [12]. This shows experimentally that the polarization fields are absent in our c-InGaN/GaN MQWs.

The bandgap energy of cubic III-nitrides is about 200 meV smaller than that of their hexagonal counterparts. Thus, for green or even longer wavelength light emission, about 10% less In is necessary to be incorporated in c-InGaN QWs than it would be with hexagonal III-nitride based devices. Furthermore, due to the absence of polarization field in cubic InGaN quantum wells, the quantum well thickness can be tuned in a wider range without degradation of the recombination efficiency. We are able to obtain strong 520 nm green PL emission from c-In$_x$Ga$_{1-x}$N/GaN (x=0.160) MQWs with 7.8 nm thick wells (see Fig. 3). The width of the emission band is comparable to that of h-InGaN/GaN MQW grown on non-polar planes emitting in the same wavelength range [13].

In summary, we have shown that by optimizing the c-InGaN growth conditions, high quality c-InGaN/GaN MQWs have been grown on 3C-SiC substrates, which have abrupt interfaces, as was manifested by the order (up to 5$^{th}$) of superlattice peaks in X-ray diffraction profile. The room temperature PL emission from 6-period c-In$_{0.16}$Ga$_{0.84}$N/GaN MQWs has a FWHM of about 240 meV and a peak wavelength of 520 nm in the green spectral region. The effective diffusion length of excess carriers in these QWs is about 17 nm. The measured integrated PL intensity of c-InGaN/GaN MQWs increases monotonically with the well thickness from 2 nm to about 8 nm which proves that polarization fields do not exist in cubic InGaN/GaN quantum wells.

**Acknowledgements**

We would like to thank H. Nagasawa and M. Abe from SiC Development Center, HOYA Corporation, for supplying the 3C-SiC substrates.

# References

**\*** present adress: Universität Karlsruhe, Institut für Angewandte Physik, 76128 Karlsruhe, Germany

**Figure captions**

Figure 1: Experimental and simulated double-crystal X-ray diffraction profiles (omega-2 theta scans) of the symmetric (002) Bragg-reflection of a 6-period c-$In_xGa_{1-x}N$/GaN MQW (x = 0.16) on 3C-SiC (001).

Figure 2: Photoluminescence spectra of 6-period c-$In_xGa_{1-x}N$/GaN MQWs (x=0.12) measured at different temperatures. (a) 300 K. (b) 2 K.

Figure 3: Room temperature photoluminescence spectra of 6-period c-$In_{0.16}Ga_{0.84}N$/GaN MQWs with different well thickness. The well thickness is indicated in the figure.

Figure 4: Integrated room temperature photoluminescence intensity of 6-period c-$In_{0.16}Ga_{0.84}N$/GaN MQWs versus well thickness. The curve is a guide for the eye.

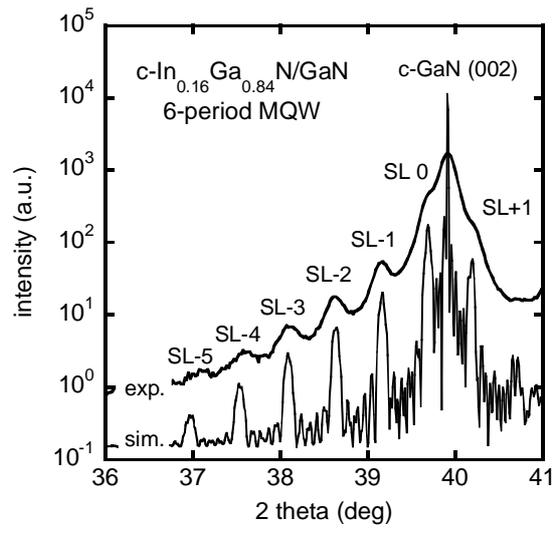

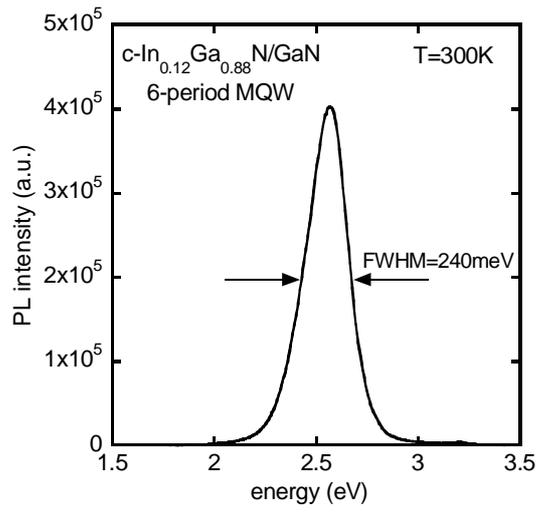 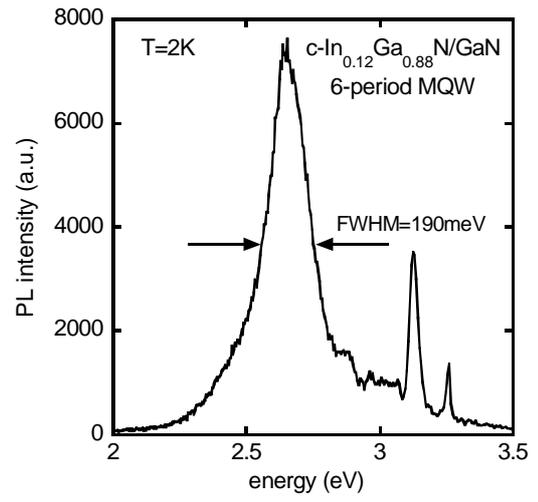

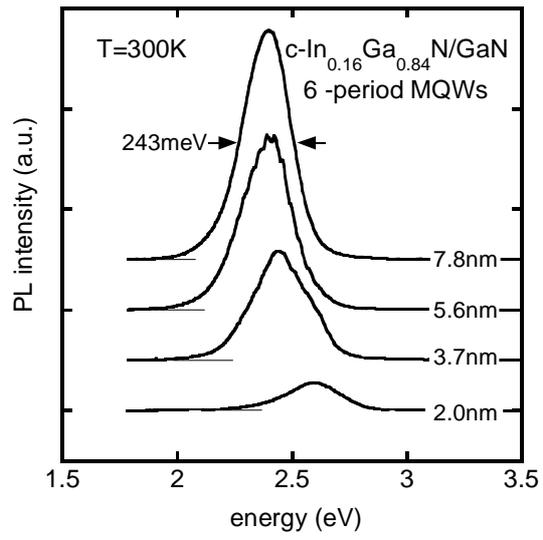

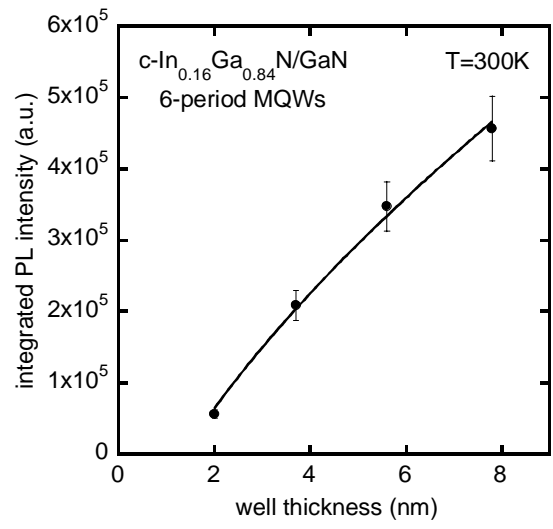